\documentclass[aps, prb,twocolumn,showpacs,preprintnumbers,amsmath,amssymb,groupedaddress]{revtex4}

\usepackage{graphicx}
\usepackage{dcolumn}
\usepackage{bm}
\usepackage{epsfig}

\begin{document}

\title{Scaling analysis of the static and dynamic critical exponents in under, over, and optimally-doped Pr$_{2-x}$Ce$_x$CuO$_{4-y}$ films}
\author{M.C. Sullivan} \email{mcsullivan@ithaca.edu}
\author{R.A. Isaacs}
\author{M.F. Salvaggio}
\author{J. Sousa}
\author{C.G. Stathis}
\author{J.B. Olson}
\affiliation{Department of Physics, Ithaca College, Ithaca, NY
14850, USA}

\date{\today}

\begin{abstract}
We report on current-voltage measurements of the zero-field
normal-superconducting phase transition in thin films of Pr$_{2-x}$Ce$_x$CuO$_{4-y}$ as a function of doping. We find that the small size of the critical regime in these materials ($\approx 25$ mK) gives rise to mean-field behavior at the phase transition with a static exponent of $\nu \approx 0.5$ for all dopings (in contrast to hole-doped  $\mathrm{YBa_{2}Cu_{3}O_{7-\delta}}$).  We also find mean-field behavior in the dynamic exponent $z$.  This indicates that Pr$_{2-x}$Ce$_x$CuO$_{4-y}$ behaves similarly to conventional superconductors in contrast to other cuprate superconductors.   However, as the transition width in our samples decreases, the dynamic critical exponent approaches $z=1.5$, similar to the critical exponent found in hole-doped $\mathrm{YBa_{2}Cu_{3}O_{7-\delta}}$.

\end{abstract}


\maketitle

\section{Introduction}

The discovery of high-temperature superconductors led to predictions
of an unusually large critical regime\cite{Lobb} as well as new
theories describing the superconducting phase transition in field and in zero
field.\cite{FFH}  These predictions led to a large number of
measurements of the phase transition in the hole-doped cuprate
superconductors via: specific heat measurements,\cite{crit}
thermal expansivity, \cite{thermex}
magnetic susceptibility,\cite{mag-suscep}
and transport.\cite{refs}

Recent attention has shifted to the phase transition of the
electron-doped superconductors R$_{2-x}$Ce$_x$CuO$_{4-y}$ (R = Nd,
Pr, La, Sm).  Previous research focused on the quantum
critical point thought to occur near optimal doping,\cite{qpt} but there has been very
little research on the more well-established vortex-glass transition
and almost none on the second-order normal-superconducting
transition in zero field.  Earlier work on
Nd$_{2-x}$Ce$_x$CuO$_{4-y}$ (NCCO) in a magnetic field reports values for the dynamic and static
critical exponents $z$ and $\nu$ for the vortex-glass transition similar to other
experiments on the hole-doped cuprates, finding $z \approx 3,4$ and $\nu
 \approx 2, 0.9$;\cite{NCCO1,NCCO2}
while work in magnetic fields from 1 mT to 1 T finds a range of exponents
$z \approx 5-9$ and $\nu \approx 0.9-1.8$.\cite{NCCO-zero} This
wide range of results is similar to results on the hole-doped
cuprates.\cite{refs}

However, the earlier works on the phase transition of NCCO were
carried out before our work proposing a more robust determination of
the critical parameters that govern the phase transition.\cite{Doug}  We have also recently uncovered
several experimental difficulties in making transport measurements
on thin films,\cite{noise, selffield} the full extent of which was
not understood when these earlier measurements were taken.  Most
notably, we have shown that the finite thickness of the films, even
of ``thick" films ($d\geq3000$~\AA)~ obscures the phase
transition.\cite{sizeeffects}  We have found that when we account for these effects, the phase transition in hole-doped $\mathrm{YBa_{2}Cu_{3}O_{7-\delta}}$ (YBCO) films and crystals yields consistent exponents for both dc and microwave conductivity measurements, $\nu = 0.68 \pm 0.07$, as predicted by three-dimensional (3D) XY theory; and $z = 1.5 \pm 0.2$,\cite{hua-su} indicating Model E dynamics.\cite{HH}

In this Article we report on the zero field normal-superconducting
phase transition of the electron-doped superconductors and present
data on the critical regime of Pr$_{2-x}$Ce$_x$CuO$_{4-y}$ (PCCO)
thin films (ranging in thickness from $d\approx2000$~\AA~ to
$d\approx3000$~\AA) for a variety of different cerium dopings ($x$). We show behavior consistent with a second-order phase transition, however, unlike recent results in YBCO, our results in PCCO are consistent with mean-field theory rather than 3D-XY theory.  Moreover, we find that at low currents, the phase transition is obscured by finite size effects, similar to results in YBCO.\cite{sizeeffects,finitesizes,yeh}

\section{Critical dynamics in Cuprates}

To understand some of the differences in the electron-doped and
hole-doped cuprates, we must look at the prediction for the size of
the critical regime, determined when mean-field theory breaks down:\cite{Lobb,goodstein}
\begin{eqnarray}
\nonumber |T-T_{c0}| &<& \frac{4 \pi \mu_o \kappa^4}{e^2 {\Phi_o}^3
H_{c2}(0)}
{k_B}^2 T_{c0}^3\\
~&<& 4.6\times 10^{-8} \frac{\kappa^4
T_{c0}^3}{H_{c2}(0)}.\label{eq:criterion2}
\end{eqnarray}
where $e$ is the base of the natural logarithm, $\kappa$ is the
ratio of the penetration depth $\lambda$ to the coherence length
$\xi$, $T_{c0}$ is the mean-field transition temperature
(measured in kelvin), and $\mu_o H_{c2}(0)$ is the GL upper critical
field (measured in tesla).\cite{errors} $H_{c2}(0)$ is not the
experimental critical field, but is the extrapolation from near the
critical regime to $T=0$.  For conventional superconductors, $\kappa \approx 10$, $T_{c0}
\approx 10$ K, and $\mu_o H_{c2}(0) \approx 1$
T.\cite{Lobb} Thus, mean-field theory breaks down only when
$|T-T_{c0}| < 1~\mu$K.  This makes the
critical regime impossible to access experimentally, and the success of mean-field theory
in describing the behavior of conventional superconductors is well documented.\cite{skocpol}

This equation can be modified to fit the anisotropic cuprate
superconductors.  In most cuprates, the $a$ and $b$ axes are nearly
identical and much smaller than the $c$ axis, so we examine the
penetration depth along the $a$ and $b$ axes ($\lambda_{ab}$)
compared to the $c$-axis ($\lambda_c$), as well as the coherence
length along the different axes ($\xi_{ab}$ and $\xi_c$).
Superconductivity occurs in the $ab$ planes.  Including anisotropy, Eq.\
\ref{eq:criterion2} becomes:\cite{FFH}
\begin{equation}
|T-T_{c0}| < 4.6\times 10^{-8} \frac{\kappa^4 T_{c0}^3}{\gamma^2
H_{c2}(0)}.\label{eq:criterion3}
\end{equation}
Here $\kappa = \lambda_{ab}/\xi_{ab}$ and the anisotropy parameter
is $\gamma = \xi_c/\xi_{ab}$.  For YBCO, using $\kappa = 100$,
$\mu_o H_{c2}(0) = 300$ T,\cite{orlando} $T_{c0}=90$
K, and $\gamma = 0.2$,\cite{FFH} we find $|T-T_{c0}| < 350$ mK.  Experimentally, researchers have found that the critical regime in YBCO can extend far beyond where GL theory breaks down and can be up to 28 times larger ($\pm10$ K).\cite{thermex}

\begin{figure}
\centerline{\epsfig{file=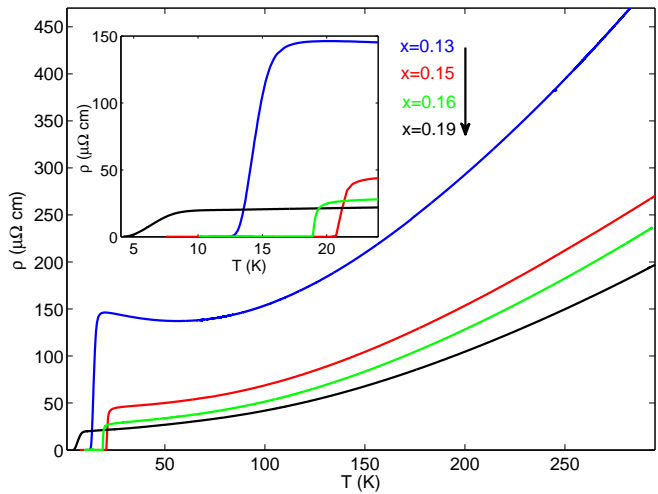,clip=,width=\linewidth}}
\caption{(Color online) Resistivity versus temperature for several dopings on $d\approx2900$~\AA~thick films patterned into $8\times80$ $\mu$m$^2$ bridges.  The inset shows the
transition region.} \label{fig:rvsT}
\end{figure}

The situation is markedly different for PCCO.  Not only is the
critical temperature reduced by roughly a factor of five, the
electron-doped cuprates are far more anisotropic. Using
$\lambda_{ab}= 2000$~\AA,\cite{PCCO-lambda} $\xi_{ab}
=80$~\AA,\cite{PCCO-xiab} $\xi_c = 3.5$~\AA,\cite{PCCO-xic} we find
$\gamma \approx 0.04$ and $\kappa \approx 25$.  With $\mu_o
H_{c2}(0) = 7$ T and $T_{c0} = 20$ K, we find that $|T-T_{c0}| < 13$
mK, more than ten times smaller than the hole-doped cuprates. This
leads to a critical regime of roughly 25 mK about $T_c$.  It is not unreasonable to expect that the critical regime in PCCO will be larger than predicted and may extend to $\pm360$ mK about $T_c$, as has been seen in YBCO.\cite{thermex}

\section{Sample Growth and experimental apparatus}

The samples for our measurements are grown via pulsed laser
deposition onto SrTiO$_3$ (100) substrates.  X-ray diffraction
verified that our films are $c$-axis orientation, and ac
susceptibility measurements show that $T_c \approx 20$ K and $\Delta
T_c\leq0.4$ K for optimal doping, indicating high quality films. Typical samples of thickness $d\approx2900$
\AA~ are shown in Fig.\ \ref{fig:rvsT},\cite{resistivitynote} showing $T_c \approx 21$ K
and $\Delta T_c\approx 0.3$ K at optimal doping ($T_c$ decreases and $\Delta T_c$ increases for under- and over-doped films).  These films are of similar
quality as most PCCO films reported in the literature.

We photolithographically pattern our film into a 4-probe bridge with
dimensions 8 $\mu$m $\times~80$ $\mu$m, etched with a low power ion
mill for 30 minutes without noticeable degradation of $R(T)$.
Contact is made to the sample leads by depositing a 200 $\mu$m thick
layer of gold on contact pads.  To ensure we do not change the
oxygen content of the sample, we do not heat the sample during the
gold deposition.

Our cryostat can achieve temperature stability  better than 1 mK at
20 K. To protect that sample from ambient magnetic fields that could
disrupt measurements,\cite{selffield} our cryostat is placed in
$\mu$-metal shields, which reduces the field at the sample to 2
$\times 10^{-7}$ T.  To reduce noise,\cite{noise} we make all connections to the probe
with shielded triax cable through low-pass $\pi$-filters.

\begin{figure}
\centerline{\epsfig{file=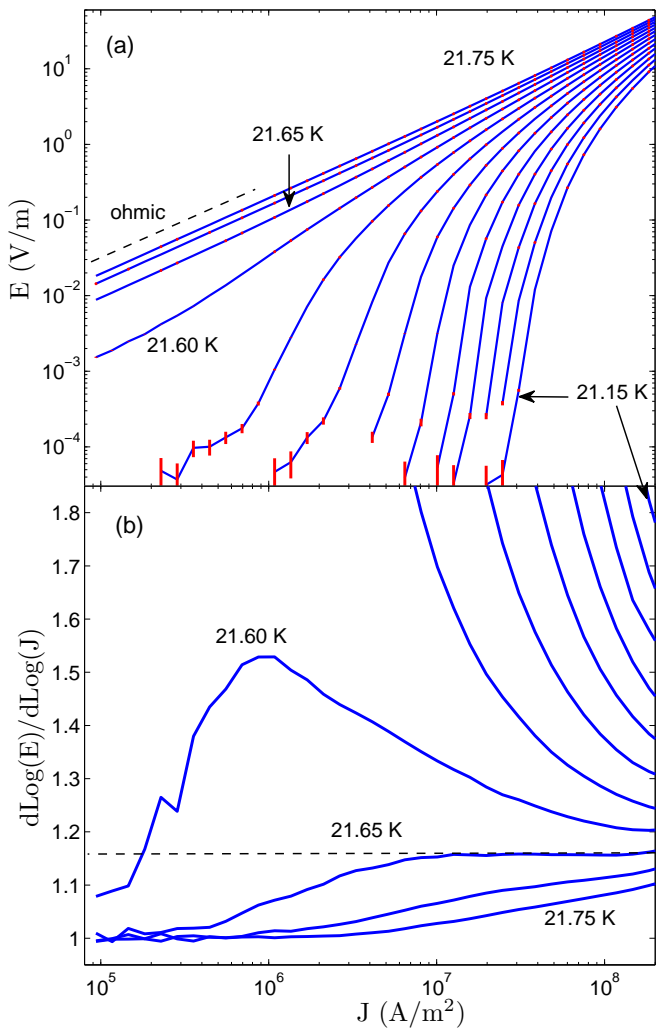,clip=,width=\linewidth}}
\caption{(Color online) (a) shows $E-J$ curves of the optimally doped film from Fig.\ \ref{fig:rvsT} in zero magnetic field (2920 \AA~ thick with bridge dimensions 8$\times$40 $\mu$m$^2$).
Isotherms are separated by 50 mK, and error bars are indicated when larger than the thickness of the lines. The dashed line indicates a slope of 1, or ohmic behavior.  Low-current ohmic tails due to finite size effects are obvious for $J \lesssim 10^6$ A/m$^2$.  (b) shows the logarithmic derivative of the $E-J$ curves.  The high current densities obey the opposite concavity criterion, indicating $T_c = 21.65$ K.  From the intercept we determine $z = 1.3 \pm 0.1$.} \label{fig:ivejs}
\end{figure}


\section{Dynamic critical exponent}

We have measured the electric field $E$ vs.\ the current density $J$ ($E-J$ curves) on 15 samples of different dopings and thicknesses varying from 1920 \AA~to 3500\AA.  The resistivities of four typical samples are shown in Fig.\ \ref{fig:rvsT}.  For every film, we measure the $E-J$ curves near the transition temperature.  A typical set of data is shown in Fig.\ \ref{fig:ivejs}(a), taken on the optimally doped film from Fig.\ \ref{fig:rvsT}.  In this plot, above $T_c$ at low
currents, we see ohmic behavior: isotherms with a slope of 1 on a
log-log plot (the dashed line).  The isotherm at 21.75 K is almost
fully ohmic.  At lower temperatures, high currents show power-law
behavior with isotherms of slope greater than 1, lower currents show ohmic behavior.  At the lowest
temperatures, the isotherms show highly non-linear behavior with
slopes approaching infinity, indicating a transition to the
superconducting state.  In this figure we can see clear evidence of a
phase transition that occurs over less than 600 mK.

Scaling analysis of the normal-superconducting phase transition predicts\cite{FFH}
\begin{equation}
E \xi^{2+z-D} / J= \chi_{\pm} ( {J\xi^{D-1}}/{T} ),
\label{eq:scaling}
\end{equation}
where $D$ is the dimension, $z$ is the dynamic critical exponent,
$\xi$ is the coherence length, and $\chi_{\pm}$ are the scaling
functions for above and below the transition temperature $T_c$.
Fluctuations are expected to have a typical size $\xi$ which
diverges near $T_c$ as $\xi \sim |T/T_c-1|^{-\nu}$, defining a
static critical exponent $\nu$.  The fluctuations are predicted to have
a lifetime $\tau$ where $\tau \sim \xi^z$, or $\tau \sim |T/T_c-1|^{-\nu z}$.

Exactly at $T_c$, the coherence length diverges while the electric field in the sample
remains finite.  From Eq.\ \ref{eq:scaling}, we can see that this is only true if, at $T_c$,\cite{FFH}
\begin{equation}
E \sim J^{\frac{z+1}{D-1}}. \label{eq:ej-tc}
\end{equation}
We have shown that a
logarithmic derivative is a sensitive tool to examine the
phase transition and find the critical temperature.\cite{Doug,sizeeffects}  From Eq.\ \ref{eq:ej-tc}, we know that the  the critical isotherm will appear as a horizontal line with
intercept $(z+1)/2$ on a logarithmic
derivative plot (assuming three dimensions).  Isotherms above and
below the critical isotherm will display opposite concavity about
the critical isotherm.  The logarithmic derivative of the $E-J$ curves in Fig.\ \ref{fig:ivejs}(a) is shown in Fig.\ \ref{fig:ivejs}(b).

Fig.\ \ref{fig:ivejs}(b) demonstrates results very similar to those
reported in YBCO, and displays the behavior predicted to occur in
the normal-superconducting phase transition. Isotherms at higher
currents  ($J > 10^6$ A/m$^2$) show the opposite concavity
criterion:\cite{Doug} isotherms above 21.65 K bend down, and
isotherms below 21.65 K bend upwards, leading us to identify $T_c =
21.65$ K as the critical temperature.

The most prominent feature of the data in Fig.\ \ref{fig:ivejs}(b) is the peak in the isotherm at 21.60 K. In fact, all power-law isotherms at or below $T_c$ (isotherms with a zero or
negative slope on the semi-log plot) eventually switch to ohmic isotherms, limited only by the sensitivity floor of our nanovoltmeter.
Power-law isotherms above $T_c$ (isotherms with a positive slope)
are expected to become ohmic at low currents, but isotherms below or
at $T_c$ \textit{are not}.

\begin{figure}
\centerline{\epsfig{file=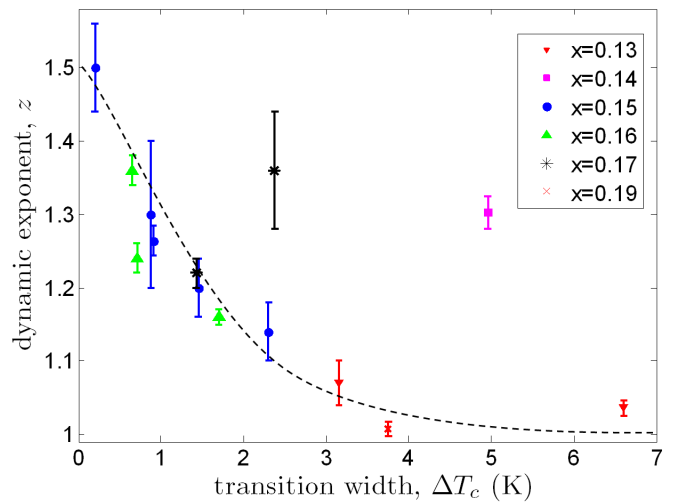,clip=,width=\linewidth}}
\caption{(Color online) Dynamic critical exponent $z$ as a function of transition width, $\Delta T_c$, with films of different dopings indicated by different symbols.  $\Delta T_c$ is measured from 90\% to 10\% of the resistance of the lowest temperature ohmic isotherm.\cite{deltaTcnote}  Results for all dopings indicate that as $\Delta T_c$ decreases, the value for the critical exponent $z$ approaches $z\approx 1.5$, similar to values found in YBCO.\cite{hua-su}  This result is easiest to see in optimally-doped ($x=0.15$) PCCO (blue dots), where a wide range of transition widths is easier to obtain.  These results indicate the critical dynamics are obscured by the width of the phase transition.  The black dotted line is a guide for the eye.} \label{fig:z}
\end{figure}

This ohmic behavior at low currents occurs due to finite size effects even in films as thick as
3000~\AA.\cite{finitesizes,sizeeffects}  Different applied currents probe
fluctuations of different sizes, leading to a minimum current
density, such that smaller current densities probe 2D fluctuations:
\begin{equation}
 J_{min} = \frac{c k_B T}{\Phi_{\mathrm{o}} d^2},
\label{eq:jmin}
\end{equation}
where $\Phi_{\mathrm{o}} = h/2e$, $d$ is the thickness of the film,
and $c$ is a constant expected to be a constant of order
one\cite{FFH,yeh} or on the order of the anisotropy
parameter $\gamma$.\cite{finitesizes}

For the 2920 \AA~film shown in Fig.\ \ref{fig:ivejs}, from Eq.\ \ref{eq:jmin}, this minimum current density occurs at $ J_{min} \approx 1.7\times10^6$ A/m$^2$ (assuming $\gamma = 1$).  We can see in Fig.\ \ref{fig:ivejs}(b) that this is indeed the current density where the isotherms below $T_c$ bend back towards ohmic behavior.  This
result indicates that, despite any differences between the
hole-doped and the electron-doped cuprates, the
normal-superconducting phase transition in both materials is
obscured in films due to finite size effects.  Conventionally, low-current ohmic tails are used to find the static critical exponent, as $E/J  \sim |T/T_c-1|^{\nu(2+z-D)}$.\cite{FFH}  Our results indicate that the low-current data in PCCO cannot be used to find the static critical exponent
$\nu$ in PCCO.

We can still use these data to find the dynamic critical exponent $z$.  This exponent is expected to be universal, but dc conductivity measurements in hole-doped cuprates in zero field have found a wide range of dynamic critical exponent values, with $z$ ranging from 1.25 to 8.3.\cite{yeh,dcmeas}  AC measurements have found both diffusive dynamics ($z=2$)\cite{zeq2} as well as Model E dynamics \cite{HH} ($z=1.5$).\cite{zeq15}  Our recent results in YBCO find $z=1.5\pm0.2$ for \textit{both} dc and ac conductivity measurements in crystals and films, when we account for finite size effects.\cite{hua-su}

To properly account for finite size effects in dc measurements, we must limit ourselves to the high current regime ($J > 10^6$ A/m$^2$) to determine the dynamic critical exponent $z$.  We recognize that the high currents obey the opposite concavity criterion, and we use the horizontal portion of the critical isotherm in Fig.\ \ref{fig:ivejs}(b)  to determine the intercept (denoted by the dotted
line).\cite{notemagsuscep}  From Eq.\ \ref{eq:ej-tc}, this intercept is equal to $(z+1)/2$, allowing us to solve
for the dynamic critical exponent, $z$.  For our data, we find
$z=1.3\pm0.1$.  A similar analysis can be conducted on all of the films of various dopings to find $z$ as a function of doping.  These values of $z$ as a function film transition width $\Delta T_c$\cite{deltaTcnote} are presented in Fig.\ \ref{fig:z}. In Fig.\ \ref{fig:z}, we find no systematic change in $z$ as a function of doping, and moreover, many of the values for $z$ are smaller than any previous measurements, and several have values of $z\approx1$, indicating ohmic behavior.  Moreover, if $z=1$, the lifetime of the fluctuations are directly proportional to the size of the fluctuations (recall $\tau \sim \xi^z$).  This ``ballistic motion" is not predicted by any theory of the phase transition.

In Fig.\ \ref{fig:z}, we see a strong correlation between the transition width $\Delta T_c$\cite{deltaTcnote} and the dynamic critical exponent.  As the transition width decreases, the dynamic exponent tends towards $z=1.5$.\cite{meaculpa}  This is easiest to see in optimally doped films ($x=0.15$).  The transition temperatures of these films are all between 19 K and 21.6 K, changing by only 12\%, but the transition widths vary from 0.2 K to 2.55 K, changing by more than a factor of 10.  Thus the variations in $z$ are driven by the transition width as opposed to the transition temperature.  Moreover, films of different dopings follow a general trend of decreasing dynamic critical exponent as transition width increases (excluding the two outliers),\cite{meaculpa} though for highly over- and under-doped films, the transition width is much wider.

From Eq.\ \ref{eq:criterion2} we predicted the size of the critical regime to be on the order of $\pm13$ mK and possibly as large as $\pm360$ mK.  This means that samples must be homogenous with transition widths smaller than 0.72 K in the best case scenario, and smaller than 0.026 K in the worst case scenario.  Wide transition widths imply sample inhomogeneity, thus the critical behavior of one part of the sample will be dominated by the larger signal generated by the mean-field behavior of different part of the sample with a slightly different $T_c$, especially as the critical regime is so small in PCCO.  This indicates that the critical regime will be obscured in films with the broad transition widths.  Our data confirm this hypothesis: At the smallest transition widths, we find $z \approx 1.5$ in optimally-doped and $x=0.16$ films, similar to results in YBCO.\cite{hua-su}  However, as the transition width increases the dynamic critical exponent tends towards $z=1$.  Our results imply that only in more homogenous films, with transition widths smaller than current transition widths will we be able to unambiguously see critical dynamics for all dopings (though whether all dopings will have the same dynamic critical exponent as optimally-doped PCCO and YBCO is unknown).  However, our films are of similar or better quality than most reported in the literature, and recent improvements in PCCO film growth\cite{patrick} remove impurities but do not decrease the transition widths.

A scaling analysis of these data indicates that $z\approx1$ for films with broad transition widths, which implies $\tau \sim \xi$ (not predicted by any theories).  However, rather than indicating a new type of phase transition, this indicates the failure of scaling analysis to describe the behavior of these films. In a conventional superconductor, there is a rapid change from the normal state with ohmic behavior ($E \sim J$) to the superconducting state, which is highly non-ohmic with $E\sim J^a$, where the power $a$ increases as current density decreases (recall that as $J$ approaches the critical current density, $a \rightarrow \infty$).  Looking only at the power $a$, for a conventional superconductor above $T_c$, we see that $a = 1$ and is independent of $J$, below $T_c$, $a(J)>1$.  This is precisely the behavior we see in our films with wide transition widths.  We identify the transition temperature as the lowest temperature isotherm whose power $a$ (the slope on the logarithmic derivative plot) does not vary with $J$.  We then use that slope to calculate $z$, recalling $a=(z+1)/2$, and if $a=1$, then $z=1$.  However, this analysis is flawed: what we are seeing is the rapid change (usually in less than 0.1 K) from ohmic behavior ($a=1$) to non-ohmic behavior ($a>1$).   In this way, the phase transition in PCCO is similar to the phase transition in conventional superconductors (rapid switch from ohmic to non-ohmic behavior), and indicates mean-field behavior rather than critical behavior.  This in turn indicates that, in regards to the normal-superconducting phase transition, PCCO usually behaves like a conventional superconductor.  Moreover, conducting a scaling analysis on this kind of system to find $z$ is intrinsically flawed -- the dynamic exponent $z$ we measure is not the critical exponent for any phase transition, but rather an indication that the critical dynamics cannot be used for films where the transition width is large compared to the critical regime.

\section{Static Critical Exponent}

We have also measured the static critical exponent $\nu$ as a function of doping.  Although we cannot use the low-current ohmic tails to measure $\nu$, as is conventionally done, in small magnetic fields\cite{FFH,hua-su}
\begin{equation}
T_c(0) - T_g(H) \sim H^{1/2\nu},\label{eq:nu}
\end{equation}
where $T_c(0)$ is the zero-field transition temperature, $T_g(H)$ is the vortex-glass transition temperature and $\nu$ is the zero-field static critical exponent.  Eq.\ \ref{eq:nu} is valid both for critical dynamics, where 3D-XY predicts $\nu \approx 0.67$, and for mean-field theory, which predicts $\nu = 0.5$.\cite{FFH}  Recent measurements on thin-film YBCO using this method found $\nu=0.68\pm0.05$.\cite{hua-su}

\begin{figure}
\centerline{\epsfig{file=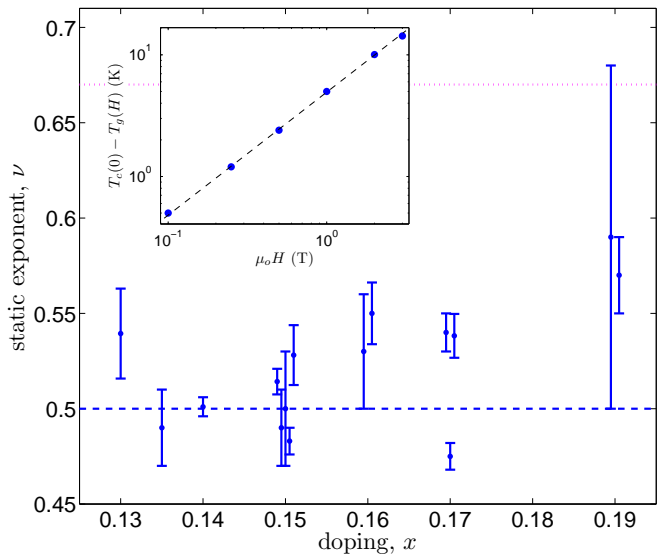,clip=,width=\linewidth}}
\caption{(Color online) Results for the static critical exponent, $\nu$, as a function of doping.  The static critical exponent was measured using the glass transition in small fields, as $T_c(0)-T_g(H) \sim H^{1/2\nu}$.  The top dashed line is $\nu = 0.67$, the result predicted from 3D-XY theory.  The lower dashed line is $\nu = 0.5$, the result predicted from mean field theory.  For all dopings, we see that $\nu \approx 0.5$, indicating a mean-field transition.   The inset shows $T_c(0)-T_g(H)$ vs.\ $H$ for the x=0.14 film; the slope of the line can be used to determine $\nu$.  A similar analysis was conducted for all data points shown here. Films with identical dopings are offset slightly along the abscissa for clarity. } \label{fig:nu}
\end{figure}

A typical measurement is shown in the inset of Fig.\ \ref{fig:nu} which shows the log-log plot of $T_c(0)-T_g(H)$ vs.\ $H$ for a film of doping $x=0.14$.  The slope of this line can be used to find $\nu$ and its error.  The results for all of the films as a function of doping are shown in Fig.\ \ref{fig:nu}.  The dashed line at the top of the figure is $\nu = 0.67$, the result predicted from 3D-XY theory.  The lower dashed line is $\nu = 0.5$, the result predicted from mean field theory.  For all dopings, we see that $\nu \approx 0.5$,  though highly over- and under-doped films deviate from $\nu \approx 0.5$.  This result is consistent with our results from measurements of the dynamic critical exponent, and indicate that the phase transition in PCCO is a mean-field transition, independent of doping.  This result is in stark contrast to recent measurements in YBCO, which show clear critical behavior, but given the smaller critical region in PCCO -- reminiscent of the small critical regimes in conventional supercondcutors, it is not surprising that PCCO behaves like a conventional superconductor in this regard.  Thus both measurements of the dynamics of these films and measurements of the static exponent $\nu$ agree and indicate that we are measuring a mean-field phase transition in PCCO.

\section{Conclusion}

We have shown that PCCO undergoes a second-order
phase transition in zero field.  The
transition obeys the opposite concavity criterion for current densities
greater than $10^6$ A/m$^2$, below these current densities,  the phase transition is
obscured by finite size effects.  We predicted the size of the critical regime in PCCO to be
$|T-T_{c0}| < 25$ mK, which means that the critical dynamics of these films
will be obscured by the large transition width in these materials ($\Delta T_c$ varies from 0.2 K to 2.55 K).  This in turn implies that the normal-superconducting phase transition in PCCO will
obey mean-field theory, similar to conventional superconductors, as opposed to critical dynamics and 3D-XY theory, as recently seen in YBCO.\cite{hua-su}  Our measurements of the static critical exponent confirm this hypothesis, as $\nu \approx 0.5$ for PCCO films of all dopings.  We also see behavior similar to conventional superconductors in the $E-J$ curves, which show a rapid change from ohmic to non-ohmic behavior.  This behavior, if analyzed using scaling analysis, gives an erroneously low value for $z$, $z \approx 1$.  However, for optimally-doped films with extremely narrow transition widths, we find that as $\Delta T_c \rightarrow 0$, $z \rightarrow 1.5$.  Thus as the transition width decreases, we are able to recover the same dynamic critical exponent found in optimally-doped YBCO.\cite{hua-su}

\begin{acknowledgements}

The authors gratefully acknowledge the assistance and support provided by R.L. Greene's lab at the Center for Nanophysics and Advanced Materials at the University of Maryland during the film growth and characterization process.
The authors also acknowledge M. Lilly for her preliminary work on this project and thank C.J. Lobb for useful discussions.  We also thank D. Tobias, S. Dutta, and R. Lewis for their help
and support of this work. We acknowledge the support of the Ithaca
College Center for Faculty Research and Development and the National
Science Foundation through Grant No. DMR-0706557.

\end{acknowledgements}


\begin{thebibliography}{99}

\bibitem{Lobb}
C.~J. Lobb, Phys. Rev. B \textbf{36}, 3930 (1987).

\bibitem{FFH}
D.~S. Fisher, M.~P.~A. Fisher, and D.~A. Huse, Phys. Rev. B
\textbf{43},  130 (1991); D.~A. Huse, D.~S. Fisher, and M.~P.~A.
Fisher, Nature \textbf{358}, 553 (1992).


\bibitem{crit}
N. Overend, M. A. Howson, and I. D. Lawrie, Phys. Rev. Lett.
\textbf{72}, 3238 (1994); G. Mozurkewich, M. B. Salamon, and S. E.
Inderhees, Phys. Rev. B \textbf{46}, 11914 (1992); S. Kamal, D. A.
Bonn, N. Goldenfeld, P. J. Hirschfeld, R. Liang, and W. N. Hardy,
Phys. Rev. Lett. \textbf{73}, 1845 (1994); A. Junod, M. Roulin, B.
Revaz, and A. Erb, Physica C, \textbf{280}, 214 (2000).



\bibitem{thermex}
V. Pasler, P. Schweiss, C. Meingast, B. Obst, H. W\"{u}hl, A. I.
Rykov, and S. Tajima, Phys. Rev. Lett. \textbf{81}, 1094 (1998).


\bibitem{mag-suscep}
K. D. Osborn, D. J. Van Harlingen, V. Aji, N. Goldenfeld, S. Oh, and
J. N. Eckstein, Phys. Rev. B \textbf{68} 144516 (2003).

\bibitem{refs} See the
references in Ref. \onlinecite{Doug} and \onlinecite{selffield} for a short list of recent transport
measurements on hole-doped cuprates and their results.


\bibitem{qpt}  See S. Sachdev, Rev. Mod. Phys. 75, 913 (2003); A. Biswas, P. Fournier, M. M. Qazilbash,
V. N. Smolyaninova, H. Balci, and R. L. Greene Phys. Rev. Lett.
\textbf{88}, 207004 (2002); Y. Dagan, M. M. Qazilbash, C. P. Hill,
V. N. Kulkarni, and R. L. Greene, Phys. Rev. Lett. \textbf{92},
167001 (2004); and references contained therein.

\bibitem{NCCO1} N.-C. Yeh, W. Jiang, and D. S. Reed, A. Gupta and F. Holtzberg,
A. Kussmaul, Phys. Rev. B \textbf{45}, 5710 (1992)

\bibitem{NCCO2}  O. M. Stoll, A. Wehner, R. P. Huebener M. Naito, Physica C
\textbf{363}, 31-40 (2001).

\bibitem{NCCO-zero} J. M. Roberts,
Brandon Brown, J. Tate, X. X. Xi, and S. N. Mao, Phys. Rev. B
\textbf{51}, 15281 (1995);


\bibitem{Doug}
D.~R. Strachan, M.~C. Sullivan, P. Fournier, S.~P. Pai, T.
Venkatesan and C.~J. Lobb, Phys. Rev. Lett. \textbf{87}, 067007
(2001); D. R. Strachan, M. C. Sullivan, and C. J. Lobb, Proc. SPIE
\textbf{4811}, 65-77 (2002).

\bibitem{selffield}
M. C. Sullivan, D. R. Strachan, T. Frederiksen, R. A. Ott, and C. J.
Lobb, Phys. Rev. B \textbf{72}, 092507 (2005).

\bibitem{noise}
M. C. Sullivan, T. Frederiksen, J. M. Repaci, D. R. Strachan, R. A.
Ott, and C. J. Lobb Phys. Rev. B \textbf{70}, 140503(R) (2004).

\bibitem{sizeeffects}
M. C. Sullivan, D. R. Strachan, T. Frederiksen, R. A. Ott, M. Lilly,
and C. J. Lobb, Phys. Rev. B \textbf{69}, 214524 (2004).

\bibitem{hua-su}
H. Xu, S. Li, Steven M. Anlage, C. J. Lobb, M. C. Sullivan, Kouji Segawa, and Yoichi Ando, Phys. Rev. B \textbf{80}, 104518 (2009).

\bibitem{HH}
P. C. Hohenberg, B. I. Halperin, Rev. Mod. Phys,\textbf{49}, 435
(1977)

\bibitem{finitesizes}
P.J.M W\"{o}ltgens, C. Dekker, R.H. Koch, B.W. Hussey, and A. Gupta,
Phys. Rev. B \textbf{52}, 4536 (1995); C. Dekker, R. H. Koch, B. Oh,
and A. Gupta, Physica C 185, 1799 (1991).

\bibitem{yeh}
N.-C. Yeh, W. Jiang, D. S. Reed, U. Kriplani and F. Holtzberg, Phys.
Rev. B \textbf{47}, 6146 (1993).

\bibitem{goodstein}
D. L. Goodstein, \textit{States of Matter}, (Dover, Mineola, NY
2002).

\bibitem{errors}
This equation differs from Lobb's formula in Ref.\ \onlinecite{Lobb}
by a factor of ${4}/{e^2}$ (roughly $\frac{1}{2}$). Lobb assumes
equipartition for one mode (a real order parameter).  We have used
equipartition for two modes (a complex order paramter), leading to a
factor of 4. Lobb also uses the small-$r$ approximation of the
correlation function ($\Gamma(r) \sim e^{-r/\xi}/r$) when $\Gamma(r)
\sim 1/r$. Because we are interested when $r \approx \xi$, we have
used $\Gamma(\xi) \sim 1/e\xi$, leading to a factor of $1/e^2$.



\bibitem{skocpol}
W. J. Skocpol and M. Tinkham, Rep. Prog. Phys., \textbf{38}, 1049-1097 (1975).




\bibitem{orlando}
T. P. Orlando and K. A. Delin, \textit{Foundations of Applied
Superconductivity}, (Addison-Wesley, 1990).


\bibitem{PCCO-lambda}
B. N. Basov and T. Timusk, Rev. Mod. Phys. \textbf{77}, 721 (2005);
A. Zimmers, R. P. S. M. Lobo, N. Bontemps ,C. C. Homes, M. C. Barr,
Y. Dagan, and R. L. Greene, Phys. Rev. B \textbf{70}, 132502 (2004);
John A. Skinta, Thomas R. Lemberger, T. Greibe and M. Naito, Phys.
Rev. Lett. \textbf{88}, 207003 (2002); J. D. Kokales, Patrick
Fournier, Lucia V. Mercaldo, Vladimir V. Talanov, Richard L. Greene,
and Steven M. Anlage, Phys. Rev. Lett. 85, 3696 (2000).

\bibitem{PCCO-xiab}
Dong Ho Wu, Jian Mao, S. N. Mao, J. L. Peng, X. X. Xi, T.
Venkatesan, R. L. Greene, and Steven M. Anlage, Phys. Rev. Lett.
\textbf{70}, 85 (1993); F. Gollnik and M. Naito, Phys. Rev. B
\textbf{58}, 11734 (1998).

\bibitem{resitivitynote}
Meausrements in Fig.\ \ref{fig:rvsT} were taken with an applied current density of $J \approx 10^6$ A/m$^2$, thus the resistive transition width are wider than those reported in Fig.\ \ref{fig:z} by roughly 40\% due to finite size effects.

\bibitem{PCCO-xic}
Y. Hidaka and M. Suzuki, Nature \textbf{338}, 635 (1989).


\bibitem{dcmeas}
J. M. Roberts, Brandon Brown, B. A. Hermann, and J. Tate,
\emph{Phys. Rev. B} \textbf{49}, 6890 (1994);
T. Nojima, T. Ishida, and Y. Kuwasawa, Czech. J. Phys. 46, Suppl.
S3, 1713 (1996); P. Voss-de Haan, G. Jakob and H. Adrian, \emph{Phys. Rev. B} {\bf
60}, 12443 (1999); K. Moloni, M. Friesen, S. Li, V. Souw, P. Metcalf, L. Hou, and M.
McElfresh, \emph{Phys. Rev. Lett.} \textbf{78,} 3173 (1997).

\bibitem{zeq2}
K. D. Osborn, D. J. Van Harlingen, V. Aji, N. Goldenfeld, S. Oh, and
J. N. Eckstein, Phys. Rev. B \textbf{68}, 144516 (2003); S. H. Han,
Yu. Eltsev, and \"{O}. Rapp, J. Low Temp. Phys. \textbf{117}, 1259
(1999).

\bibitem{zeq15}
F. S. Nogueira, D. Manske, Phys. Rev. B \textbf{72}, 014541 (2005);
J. T. Kim, N. Goldenfeld, J. Giapintzakis, and D. M. Ginsberg, Phys.
Rev. B \textbf{56}, 118 (1997); Vivek Aji and Nigel Goldenfeld,
Phys. Rev. Lett. \textbf{87}, 197003 (2001).


\bibitem{patrick}
G. Roberge, S. Charpentier, S. Godin-Proulx, P. Rauwel, K.D. Truong, P. Fournier, J. Crystal Growth, \textbf{311}, 1340-1345 (2009).

\bibitem{notemagsuscep}
The $T_c$ as measured via the opposite concavity criterion here and the magnetic susceptibility agree to within the measurement error of the thermometers ($\pm 0.25$ K), indicating that the $T_c$ as measured via the opposite concavity criterion agrees with the bulk measurement of the transition temperature.

\bibitem{deltaTcnote} We calculated $\Delta T_c$ by using the 90\% - 10\% criterion; i.e., we found the resistivity of lowest-temperature ohmic isotherm from the $E-J$ curves, and then found the temperatures that corresponded to 90\% of that resistivity and 10\% of that resitivity to calculate $\Delta T_c$.  We used the resisitivity at $J = 10^8$ A/m$^2$ to avoid finite-size effects.   An alternative method of measuring $\Delta T_c$ using magnetic susceptibility yields similar results (with a smaller scale along the abscissa).

\bibitem{meaculpa} There are two films ($x=0.14$ and $x=0.17$) that deviate strongly from the general trend shown in Fig.\ \ref{fig:z}.  At this time, we are unsure why these films deviate from the general trend, as growth, procecssing, and measurement were the same for these films as for all the others presented in Fig.\ \ref{fig:z}.  However, we note that others film of the same doping and similar dopings do follow the trend, and so our overall conclusion is still valid.



\end{thebibliography}
\end{document}